\documentclass[prd,nofootinbib,twocolumn,showpacs]{revtex4}

\usepackage{graphics}

\begin{document}

\tighten

\title {The Creation of Defects with Core Condensation}

\author{Nuno\ D.\ Antunes}  \affiliation{
Center for Theoretical Physics, University of Sussex, \\
Falmer, Brighton BN1 9WJ, U.K.}
\author{Pedro\ Gandra} \author{Ray\ J.\ Rivers}
\author{A.\ Swarup}
\affiliation{ Blackett Laboratory, Imperial College\\
London SW7 2BZ, U.K.}

\begin{abstract}
Defects in superfluid $^3He$, high-$T_c$ superconductors, QCD
colour superfluids and cosmic vortons can possess
(anti)ferromagnetic cores, and their generalisations. In each case
there is a second order parameter whose value is zero in the bulk
which does not vanish in the core. We examine the production of
defects in the simplest 1+1 dimensional scalar theory in which a
second order parameter can take non-zero values in a defect core.
We study in detail the effects of core condensation on the defect
production mechanism.

\end{abstract}

\pacs{03.70.+k,  05.70.Fh, 03.65.Yz}

\maketitle

\section{Introduction}
\label{intro}

Since phase transitions take place in a finite time, causality
guarantees that correlation lengths remain finite. If the symmetry
breaking permits non-trivial homotopy groups the frustration of
the order parameter fields is resolved by the creation of
topological defects to mediate between the different ground
states. Since defects are, in principle, observable, they provide
an excellent experimental tool for determining how transitions
occur.

Zurek\cite {zurek1,zurek2} and Kibble\cite{kibble} originally
suggested that causality alone is sufficient to bound the initial
density of defects arising in a continuous transition, whether in
condensed matter or quantum field theory. The analysis is very
general, and depends on the fact that there is a maximum speed
(e.g. the speed of light or the speed of sound) at which the
system can become ordered. The ZK causal arguments can be
quantified in many variants, and we will not rehearse them here.
It is sufficient to consider a system with critical temperature
$T_{c}$, cooled through that temperature at a rate
\begin{equation}
\frac{1}{T_c}\frac{dT}{dt}\bigg\vert_{T_c} = -\frac{1}{\tau_Q},
\label{rate}
\end{equation}
thereby defining  the quench time $\tau _{Q}$. The prediction is
that, if ${\bar{\xi}}$ is the defect separation at the time of
defect formation, then
\begin{equation}
{\bar{\xi}}= f\xi _{0}\bigg(\frac{\tau _{Q}}{\tau _{0}%
}\bigg)^{\sigma }\gg \xi _{0},  \label{xibar}
\end{equation}
where $\tau_0$ is the relaxation time for short wavelength modes
and $\xi_0$, also determined from the microscopic details of the
system, characterises the size of cold defects. The coefficient
$f$ is an undetermined efficiency factor, taken to be $O(1)$, but
anticipated to be greater than unity.

We term $\sigma $ the Zurek-Kibble (ZK) characteristic index. Its
mean-field values are typically $\sigma = 1/3$ for relativistic
systems and $\sigma = 1/4$ for strongly damped non-relativistic
systems. Experiments on a range of condensed matter systems
(superfluid $^3He$ \cite{grenoble,helsinki}, high-$T_c$
superconductors \cite{technion1,technion2}, Josephson tunnel
junctions (JTJs) \cite{Monaco1,Monaco2}) give results that are
commensurate with (\ref{xibar}).

This may seem paradoxical in that subsequent analytic
approximations \cite{karra,ray2} and numerical simulations
\cite{laguna,antunes,moro} have shown that, for simple systems,
the scaling behaviour of (\ref{xibar}) is understood, not so much
in terms of causal bounds, but in terms of the instabilities of
the time-dependent Ginzburg-Landau (TDGL) theory, whose
dissipative behaviour controls which regime we are in. However,
where scaling is appropriate it is found that causality arguments
identify the correct engineering dimensions for the scaling
behaviour of defect densities \cite{ray2,moro} and (\ref{xibar})
survives \cite{us2}.

For symmetry breaking of local gauge theories the result
(\ref{xibar}) is not complete because of the freezing in of long
wavelength modes of the gauge fields \cite{rajantie}. However, for
the high-$T_c$ experiments of \cite{technion1,technion2} and the
JTJs of \cite{Monaco1,Monaco2}) the effect of this additional
mechanism is small and we shall ignore it, given that it does not
occur for the simple model that we shall solve numerically later.

For each of the condensed matter systems listed above the relevant
topological defects are vortex lines (strings).  Again, in the
context of the early universe, strings (cosmic strings) are the
almost inevitable consequence of symmetry breaking in the most
obvious extensions of the standard model for electroweak
unification \cite{mairi}. Observation has suggested the possible
existence of cosmic strings \cite{saz1,saz2,schild} but, as yet,
the evidence is not compelling.

Most simply, for both condensed matter and the early universe, the
most studied and best understood strings arise from the breaking
of a global or local $U(1)$ symmetry, in which the order parameter
can be represented by a single complex scalar field. In such cases
the structure of the vortex is quite simple, with a trivial core,
in the centre of which the order parameter vanishes, restoring the
$U(1)$ symmetry there. That is, a string is a simple tube of false
vacuum or ground-state, trapping flux if the symmetry is local.
Although this is appropriate for the phases of the condensed
matter systems for which experiments have been performed, it is a
simplification since both superfluid $^3He$ and high-$T_c$
superconductors have order parameters $\vec\Phi$ with several
components. The unstable ground state ('false vacuum') corresponds
to $\vec\Phi = 0$.  We can separate the components of $\vec\Phi$
into two types $\vec\Phi = (\vec\phi ,\vec\eta )$ so that, in
bulk, the $\vec\phi$ fields condense (i.e. are non-zero), whereas
the $\vec\eta$ fields do not. For any defect this characterises
the situation in its exterior and, for a 'normal' defect,  {\it
both} $\vec\phi$ and $\vec\eta$ are zero in the core. However, in
some parts of parameter space it may require less energy to
produce defects with 'abnormal' cores, in which $\vec\phi = 0$,
but the $\vec\eta\neq 0$. The existence of defects inside which
the $\vec\eta$ fields condense is not a question of topology,
merely one of energetics.

This is the case for superfluid $^3He-B$ with its nine complex
order parameters, for which there is experimental evidence for
superfluid $^3He-B$ vortices with ferromagnetic $^3He-A$
cores\cite{Volovik1,Volovik2}. Also, this phenomenon has been
predicted to occur in the idealised SO(5) model
\cite{Zhang,SO5Vortons} of high-temperature superconductivity,
where, for appropriate doping, the cores of the conventional
Abrikosov vortices are antiferromagnetic. There has been recent
experimental evidence to support this theoretical
picture\cite{vaknin,lake,dai,mitrovic,miller}.

This is not just a phenomenon of condensed matter physics. It has
been argued \cite{Kaplan,KStrings, KVortons} that such strings
also occur in the so-called colour superconducting phase of QCD,
that is believed to be realized when the baryon density is a few
times larger than nuclear density \cite{CFL}, and hence in neutron
stars. In fact, that such a phenomenon can occur in relativistic
systems was first discussed by Witten \cite{Witten} on looking at
generalisations of cosmic strings. His model is the simplest of
all that is compatible with early universe cosmology: consider a
two component system described by two complex scalar fields
$(\phi,\eta)$ with an approximate
  $U(2)$ symmetry. If the $U(2)$ symmetry between fields $\phi$ and
$\eta$ is explicitly  broken down to $U(1) \times U(1)$, the
$\phi$ condensation might be energetically more favorable than
$\eta$ condensation, and the ground state will be given by
$\langle\phi\rangle \neq 0$ and $\langle\eta\rangle = 0$. As
intimated above, this system permits $\phi$ vortices characterized
by the phase of $\phi$ field varying by an integer multiple of $2
\pi$ as one traverses a contour around the vortex core. Witten
showed that, if the approximate $U(2)$ symmetry is broken only
weakly, the $\eta$ field may condense inside the core of the
$\phi$ string, breaking the corresponding $U(1)$ symmetry in the
core.

One consequence of this 'core-condensation' is that \cite{Turner,
Hindmarsh,Davis, DavisShell} it provides a way to stabilize a
string loop against shrinking, as would normally be the case
because of the string tension. The condensed matter counterpart to
Witten's model of approximate $U(2)$ symmetry breaking is in
two-component BECs \cite{met}, for which both core condensation
and this stabilisation may already have been seen.

Most of the work cited above has addressed the static properties
of defects with core condensation. In this paper we are interested
in the dynamics of the creation of such defects. Insofar as the
scaling behaviour of (\ref{xibar}) is truly a consequence of
causality we would expect it to be equally true for defects with
and without core condensation. However, it may not be so simple in
that one consequence of core condensation is that the interactions
between vortices can be drastically altered by the presence of
non-trivial cores \cite{MacKenzie}. In the following sections we
shall consider the simplest model permitting core condensation, to
check whether the formation accords with the simple Kibble-Zurek
scaling laws in $\tau_Q$ of (\ref{xibar}).

\section{The $O(2)\rightarrow Z_2\times Z_2\rightarrow Z_2$ model
in 1D}

Our model is a simplified version of more realistic dissipative
systems, such as the Ginzburg-Landau description \cite{Zhang2} of
high-$T_c$ superconductors. However, with the early universe in
mind we extend it to a relativistic underdamped model when
appropriate.

Specifically, let us consider the symmetry breaking
$O(2)\stackrel{ESB} \rightarrow Z_2\times
Z_2\stackrel{SSB}\rightarrow Z_2$ in one dimension, where $ESB$
denotes explicit symmetry breaking by the introduction of terms in
the action and $SSB$ denotes spontaneous symmetry breaking. We
consider a free energy of the form
\begin{equation}
F = \int dx\bigg[{1\over 2}[(\partial_x\phi)^2 + (\partial_x{
{\eta}})^2] + V(\phi,{\eta}) \bigg], \label{F(1,1)}
\label{eq:seven}
\end{equation}
where
\begin{equation}
V(\phi,{\eta}) = {1\over 2}a(T)\phi^2 + {1\over 2}a(T)\beta\eta^2
+ {b\over 4}(\phi^2 + {\eta}^2)^2+\frac{a^2}{4b}.
\label{potential}
\end{equation}
for which we have chosen to implement the 'phase transition'
explicitly through the Landau form $a(T) = -a'(1-T/T_c)$, where
$a'>0$. The fact that there is no true transition in one dimension
is irrelevant for the rapid changes in temperature that we shall
consider \cite{laguna}.

When $\beta = 1$ $F$ possesses $O(2)$ symmetry, and we can take
$\phi$ and $\eta$ to be the real and imaginary parts of a complex
field $\Phi$. There is a minor complication in that, if we fix the
phase of $\Phi$ to be zero, say, at infinity, making space $S^1$,
the winding number of the phase is a conserved charge. However,
the configurations with non-zero charge are Skyrmions, and not the
defects of relevance to the ZK scenario, of which none exist in
the absence of topological charge when there are no fixed boundary
conditions.

The $O(2)$ is broken explicitly to $Z_2\times Z_2$ when $\beta\neq
1$, which will be the case of interest for defect formation.  In
1D, the defects are kinks. For example, to convert $F$ of
(\ref{F(1,1)}) into an $SO(5)$ model for high-$T_c$ along the
lines of \cite{Zhang2} we elevate $\phi$ into the two-component
complex field that couples to the electromagnetic field, and
elevate $\eta$ to the three-component N\'{e}el vector of
antiferromagnetism. A similar transmutation would recreate a model
like Witten's. In this regard, we note that the potentials
presented in the literature are not quite identical in the way the
different sectors couple. They only become so on imposing
non-linear constraints (e.g. $\vec\phi^2 + \vec\eta^2 = f^2$), as
in \cite{Zhang2}. At an effective level this can make little
difference to the statics of defects (e.g. \cite{arovas}), but for
the dynamics we revert to the more fundamental linear sigma model
of (\ref{F(1,1)}), modelled on \cite{arovas} and \cite{alama}.
Phase-transitions leading to kinks and walls with core condensation
 and displaying  repulsive interactions
were considered in the context of a $S_5\times Z_2\rightarrow S_3 \times S_2$
theory in \cite{vachaspati}. In that case the authors were 
concerned mostly with the long-time behaviour of the system after the
transition.
In particular, significant changes to the scaling beaviour of
domain wall networks were observed as well as the formation of defect
 lattices as the final product of the evolution. Here we will concentrate 
on the period straight
after the transition, namely on the process of defect formation, rather than
on the long-time dynamics.

The properties of the translation
 invariant configurations that minimize the free-energy
above depend on the value of $\beta$. Defining $\phi_0^2=-a/b$ and
$\eta_0^2=-\beta a/b$ we have that, for $\beta>1$,
\begin{eqnarray}
F\left[\phi_{0},0\right]>F\left[0,\eta_{0}\right], \qquad \beta > 1%
\label{eq:eight}
\end{eqnarray}
corresponding to an ground state with $\phi(x)=0$.

 For
$0<\beta<1$ the opposite relation holds
\begin{eqnarray}
F\left[\phi_{0},0\right]<F\left[0,\eta_{0}\right], \qquad 0<\beta < 1%
\label{eq:nine}
\end{eqnarray}
and the preferred ground state configuration is now
$\eta(x)=0$, $\phi^2(x)=\phi_0^2$. For $\beta<0$, the case corresponding
to Eq.~(\ref{eq:eight})
 holds again, as $\beta$ is
now negative and $\eta$ is therefore constrained to be zero.

\subsection{Kink solutions}

We will concentrate on the $0<\beta<1$ regime and consider
kink-like solutions that minimize the action in
Eq.~(\ref{eq:seven}) for the set of boundary conditions
$\phi(-\infty)=-\phi_0, \eta(-\infty)=0$ and
$\phi(+\infty)=\phi_0, \eta(+\infty)=0 $. For any value of $\beta$
in this region the free-energy is always stationary for the
straightforward kink solution obtained by setting
\begin{equation}
\phi (x) = \phi_0 \tanh (l_k x),\,\,\,\, \eta(x) =0.
\label{old_kink}
\end{equation}
where $l_k =(-a/2)^{1/2}$ is the kink size. The free-energy of
this configuration is $\beta$ independent and given by
\begin{equation}
F_k =\frac {2\sqrt{2}}{3}b^{1/2}\phi_{0}^{3}%
\label{eq:fourtythree}
\end{equation}

\begin{figure}
\scalebox{0.45}{\includegraphics{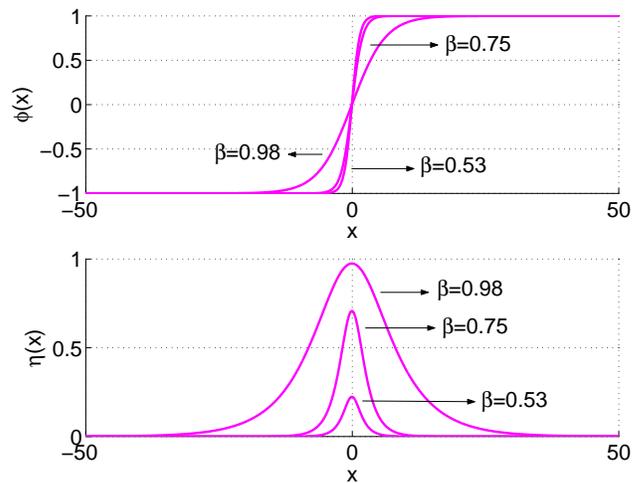}}
\caption{Profiles of $\phi(x)$ and $\eta(x)$ for several values of $\beta>0.5$
and $a=-1.0$ and $b=1.0$}
\label{profiles}
\end{figure}
It is possible that for certain values of $\beta$ other stable
configurations exist
 with a lower free-energy \cite{arovas,alama}.
It turns out that, for this particular system, we were able to
find such solutions analytically.

Any static solution  obeys the time-independent equations of
motion derived from the free-energy Eq.~(\ref{eq:seven}):
\begin{eqnarray}
\partial_x^2\phi&=&a\phi+b(\phi^2+\eta^2)\phi \label{eq_phi} \\
\partial_x^2\eta&=&\beta a \eta + b(\phi^2+\eta^2)\eta. \label{eq_eta}
\end{eqnarray}
We start by assuming that $\phi(x)$ has the usual kink-profile
\begin{equation}
\phi (x) = \phi_0 \tanh (x/l), \label{phikink}
\end{equation}
 where $l$, the size of the core
can now depend on $\beta$. The trivial kink profile with
$l_k=\sqrt{-2/a}$ and $\eta=0$ is obviously a solution for all
values of $\beta$. In order to find other possible solutions, we
assume $\eta(x)\neq 0$ and replace the general kink profile (with
arbitrary core size $l$)
 into the equation for $\phi$, Eq.~(\ref{eq_phi}).
Solving in terms of $\eta$ we obtain:
\begin{equation}
\eta(x)=\phi_0 \left(1+\frac{2}{a l^2}\right)^{1/2}
\frac{1}{\cosh(x/l)}. \label{etakink}
\end{equation}
Demanding that this result satisfies the second equation, Eq.~(\ref{eq_eta})
we obtain an explicit expression for the kink core size
\begin{equation}
l=\frac{1}{\sqrt{-a}}\frac{1}{\sqrt{1-\beta}}=\frac{l_k}{\sqrt{2-2\beta}}.
\label{core_size}
\end{equation}

The full solution is then given by
\begin{eqnarray}
\phi (x) &=& \phi_0 \tanh (x/l)\\
\eta(x) &=&\pm \phi_0 \sqrt{2 \beta-1} \frac{1}{\cosh(x/l)}.
\end{eqnarray}
Note that we have included the two possible signs for the $\eta$
field. In general, for each choice of $\phi$ profile, there will
be two allowed symmetric configurations for $\eta$. We will refer
to solutions for which $\phi$ interpolates between $-\phi_0$ and
$\phi_0$ as $x$ goes from $-\infty$ to $\infty$ as kinks. These
can be positively of negatively charged, depending on whether
$\eta$ is positive or negative. Conversely, anti-kinks will have
$\phi(-\infty)=\phi_0$ and $\phi(\infty)=-\phi_0$. We will say an
anti-kink has negative charge if $\eta>0$ and positive charge
otherwise. This convention for the charge signs for kinks and
anti-kinks corresponds to the winding direction of the solution in
the $(\phi,\eta)$ plane, mimicking the $U(1)$ charge definition in
the $\beta\rightarrow 1$ limit. As we will see below, the
interaction potential between kinks and anti-kinks will depend on
their charges, as defined above.

The free-energy of the non-trivial kink profile can
 be evaluated explicitly, being given by
\begin{equation}
F=F_k \sqrt{2-2\beta}\left(\frac{1}{2} + \beta\right),
\end{equation}
where $F_k$ the energy of the $\eta=0$ solution as defined in
 Eq.~(\ref{eq:fourtythree}).
Such a kink  is rather like  a a slice through a configuration
known as a 'dark-bright vector soliton'  in two-component BEC
\cite{Kinks}, which consists of a domain wall formed by one
condensate with the second condensate confined to the wall's
centre. The difference lies in the fact that the BEC condensate
obeys non-dissipative Gross-Pitaevskii equations, rather than the
(relativistic) dissipative TDGL equation appropriate to condensed
matter and the early universe.

\begin{figure}
\scalebox{0.45}{\includegraphics{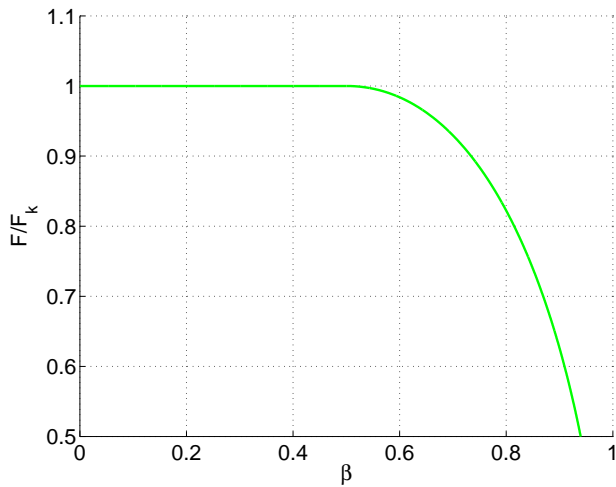}}
\caption{The free energy of the most stable kink configuration against $\beta$,
 normalized by the energy of the $\eta=0$ kink.}
\label{free-energy}
\end{figure}

\begin{figure}
\scalebox{0.45}{\includegraphics{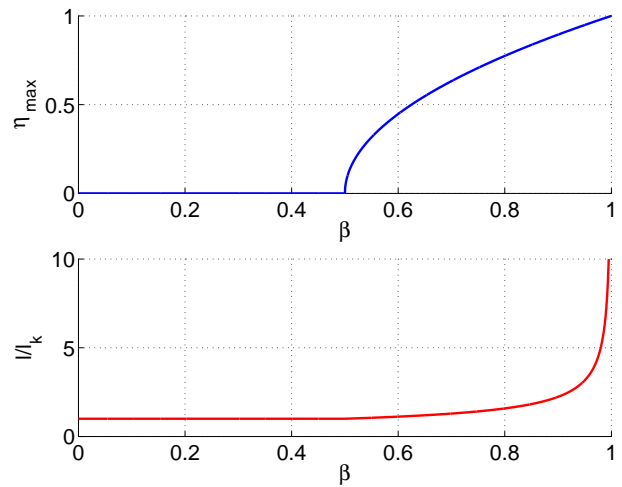}}
\caption{Maximum value of $\eta(x)$ and core-length of the most stable configuration  versus $\beta$.}
\label{eta_l}
\end{figure}

 The above expressions for the fields' profiles and the free-energy
of the non-trivial solution are defined only for $\beta>\beta_c$,
where $\beta_c=0.5$, a value independent of the parameters of the
theory. For $\beta$ above the critical value we have that $F<F_k$,
the free-energy of the non-trivial solution is smaller than that
of the $\eta=0$ kink. In this region of parameter space the
configuration in Eqs.~(\ref{phikink}-\ref{etakink}) is
energetically more stable and should be more likely to form as the
outcome of a phase-transition. As $\beta\rightarrow\beta_c^{+}$
this solution converges smoothly to the trivial kink profile, with
$l \rightarrow l_k$, $\eta\rightarrow 0$ and $F\rightarrow F_k$.
For $\beta<\beta_k$, Eq.~(\ref{old_kink}) corresponds to  the
single valid solution. These points are illustrated in
Figs.~\ref{profiles}-\ref{eta_l}. In Fig.~\ref{profiles} we show
several field profiles in the $\beta>\beta_c$ regime. As $\beta$
approaches its critical value the size of the kink core decreases
tending to $l_k$, and the magnitude of the peak of $\eta$ goes to
zero. In Fig.~\ref{free-energy} we have the free energy of the
most stable configuration, for $\beta$ in the interval $(0,1)$.
Finally, in Fig.~\ref{eta_l} we show, for the most stable
configuration, the maximum value of $\eta$ and the size of the
kink core. As expected, both these and the free-energy are
constant below $\beta_c$. It is interesting to note that while all
these quantities are continuous at $\beta_c$, their derivatives
behave in different ways. In particular, $F'(\beta)$ is continuous
throughout the interval $(0,1)$, whereas the core size has finite
discontinuous derivative at $\beta_c$ and $\eta_{\rm max}$ has
diverging derivative at $\beta_c^{+}$.

\subsection{Kink/anti-kink interaction}
\label{interaction}

 The problem we want to address here is to determine what happens when
 a kink and an anti-kink are put in the vicinity of each other. For the
trivial profiles corresponding to $\beta<0.5$, it is well known
that a kink and anti-kink attract each other with a force that
decreases exponentially with the separation of the cores. In a
setting where kinks and anti-kinks form as a consequence of a
phase-transition, this allows for pair annihilation processes. In
our case, because of the presence of the extra field, kinks can
have two distinct charges and it is reasonable to expect these to
play a role in their interaction. If we imagine a kink/anti-kink
pair with the same charge in the limit $\beta\sim 1$, the result
is a configuration close to a ``springy'' $U(1)$ winding. The
$U(1)$ spring will tend naturally to ``stretch'' itself,
corresponding, in terms of the kink/anti-kink pair, to a repulsive
interaction. The opposite reasoning suggests that kinks and
anti-kinks with symmetric charges should attract each other.

 In order to check this scenario and obtain effective interaction
potentials we followed the gradient flow approximation \cite{Manton}.
This consists in setting up an initial kink/anti-kink configuration
and  evolving it numerically according to the first-order equation
of motion corresponding to the free energy in Eq.~(\ref{eq:seven}):
 \begin{eqnarray}
\partial_t \phi_a = \partial^2_x \phi_a
-\frac{\partial V}{\partial \phi_a},
\label{grad_flow}
\end{eqnarray}
where the index $a$ runs from $1$ to $2$ with $\phi_1=\phi$ and
$\phi_2=\eta$.
 This forces the solution to the be locally on a minimal energy
 configuration for all $t$, in the sense that its trajectory should follow
 the ``valleys'' of the free-energy landscape.  At given simulation times
 we define the
 defect separation $r$ as the distance between the two zeros of $\phi$ and
 we measure the total free-energy of the system. A potential $V(r)$ for this
 ``moduli'' is defined by subtracting the energy of two isolated
 kinks to the energy measured for each value of $r$ in the simulation.

\begin{figure}
\scalebox{0.45}{\includegraphics{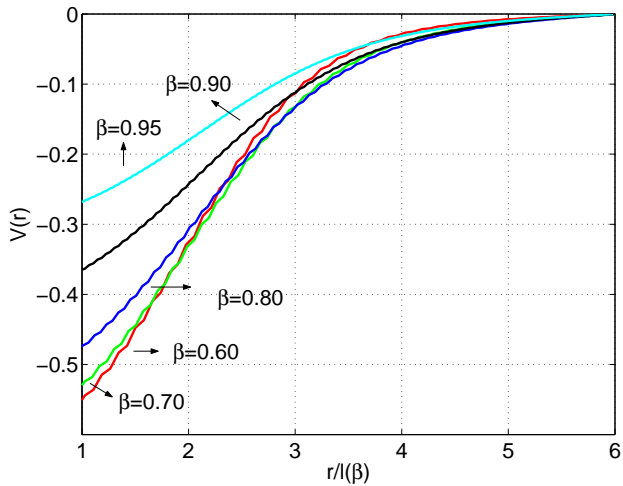}}
\caption{The gradient flow interaction potential between a kink and an
 anti-kink with opposite charges, for several choices of $\beta$. The
kink/anti-kink separation is given in units of core size.}
\label{attraction}
\end{figure}

\begin{figure}
\scalebox{0.45}{\includegraphics{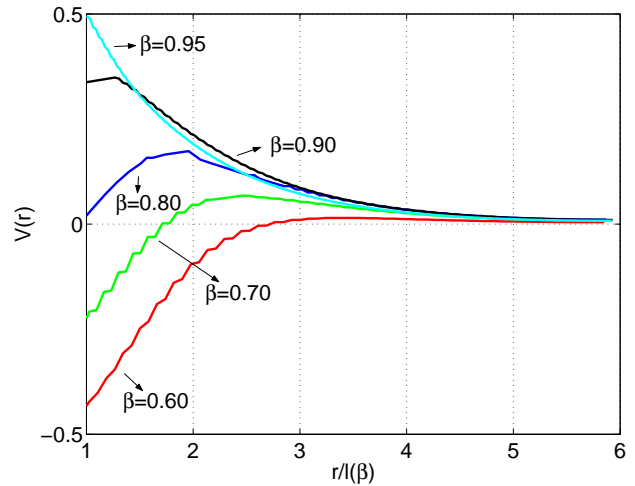}}
\caption{Interaction potential for a kink/anti-kink pair with
identical charges, for a selection of values of $\beta$. As in
Fig.~\ref{attraction}, the distance is in terms of the static
solution core size.}
\label{repulsion}
\end{figure}

 In Figs. \ref{attraction} and \ref{repulsion} we show $V(r)$ for several
 values of $\beta$, for pairs with opposite and identical charges respectively.

 As expected, the force between opposite charged kinks and anti-kinks
is always attractive. As $\beta$ increases and the kinks become
less localized, the steepness of the potential decreases
indicating a weakening of the interaction force. A kink/anti-kink
pair with identical charges displays, on the contrary, a repulsive
interaction for distances above a certain threshold. For smaller
distances the interaction seems to be always attractive, though
results below the core size, $r<l(\beta)$, should not be trusted
as the ``moduli'' becomes ill-defined. As $\beta$ increases, the
height of the potential maximum becomes larger and its location
moves towards smaller values of $r$.
 Though for values of $\beta$ just
above the $1/2$ threshold the size of the potential barrier is
likely to have negligible effects, as $\beta\rightarrow 1$ the
kinetic energy needed to surpass the barrier will become
increasingly high. As a consequence of the repulsive nature of
their interaction, same-charge kinks may be prevented from
colliding and we can expect a regime where pair annihilation will
be suppressed. As we will see, this will lead to qualitative
changes in defect production in a typical phase transition.

\section{Numerical Simulations of Defect Formation}

We now proceed to numerically evaluate the formation of these various types
 of defects by using a Langevin equation to simulate a series of quenched
transitions. Our main goal will be to check whether the types of
scaling predicted by the Kibble-Zurek scenario in (\ref{xibar})
are compatible with the presence of non-trivial defects as
described in the previous section. We will evolve the general
second-order Langevin equation
\begin{eqnarray}
\partial_t^2 \phi_a-\partial^2_x \phi_a + \alpha^2 \,\partial_t\phi_a
+\frac{\partial V}{\partial \phi_a} = \alpha \zeta_a,
\label{eom}
\end{eqnarray}
where the index $a$ runs from $1$ to $2$ with $\phi_1=\phi$ and
$\phi_2=\eta$.
 $\zeta_a$ is a gaussian noise term obeying
\begin{equation}
 \langle\zeta_a(x',t') \zeta_b(x,t)\rangle=\Gamma
\delta(x'-x) \delta(t'-t)\delta_{ab},\,\,\,\,
\langle \zeta_a(x,t) \rangle=0 .
\end{equation}
and $\alpha$ measures both the amplitude of the noise and of the
dissipative first-order time derivative term. The relation between
these two terms ensures that the fluctuation-dissipation
theorem is satisfied and guarantees that for very large times thermal
equilibrium at temperature $\Gamma/2$ is reached.
A simpler version of this model with one single field has been used
successfully in several studies of defect formation in 1+1 dimensions
\cite{laguna,us}; we refer the reader to these for more a detailed discussion
of the model and its numerical implementation.

Our goal here is to simulate a ``phase-transition'' with general
quench rate $\tau_Q^{-1}$. We start at high-temperature with $a>0$
and, after allowing the fields to equilibrate, we decrease $a(T)$
with $T$ changing according to Eq.~(\ref{rate}). As $T$ goes below
$T_c$, a series of defects forms separating regions in alternating
vacua. Finally, $a(T)$ settles to a constant negative value and
the defect population enters a period of relative slow evolution,
with occasional pairs of defects and anti-defects annihilating. At
this point, we measure the defect density $\rho = 1/\bar{\xi}$ by
counting the number of kinks and anti-kinks in the field
configuration. The defects are identified as zero crossings of the
$\phi$ field, suitably coarse-grained over a few lattice sites.
The coarse-graining is particularly relevant for high values of
$\beta$, when the size of the defect core becomes considerably
large, and small fluctuations around zero might lead to defect
over-counting.

 By repeating the above process
for several values of the quench rate we obtain $\rho(\tau_Q)$ and
fit the results to a power-law $A \tau_Q^{-\sigma}$. Note that
depending on the magnitude of the parameter $\alpha$,
Eq.~(\ref{eom}) may describe either a relativistic system (small
$\alpha$) or one where dissipation is dominant (large $\alpha$).
The former type of setting will be typical of high-energy and
cosmological systems, whereas the later will correspond to
condensed-matter systems. As discussed in Section \ref{intro} we
should obtain, according to the standard KZ predictions,
$\sigma=1/4$ for the over-damped regime and $\sigma=1/3$ for the
under-damped case.

\begin{figure}
\scalebox{0.45}{\includegraphics{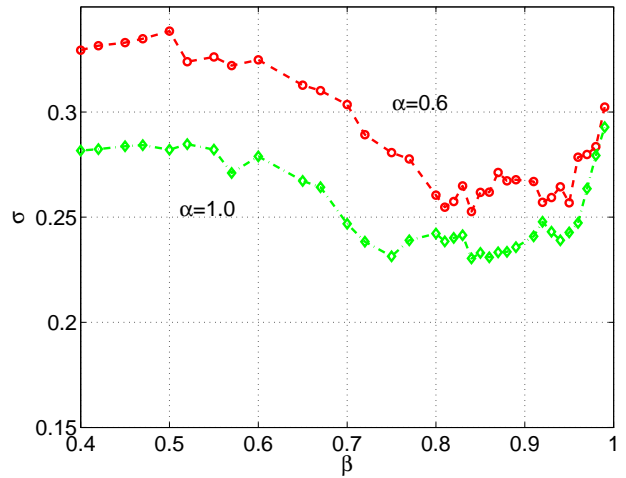}}
% {\includegraphics{sigma.eps}}
 \caption{The scaling
exponent $\sigma$ as a function of  $\beta$ for $0.4<\beta<1$, in
the under-damped ($\alpha=0.6$) and over-damped ($\alpha=1.0$)
regimes. The error-bars, not shown for clarity, are of order of
$10\%$ of the values measured.} \label{results}
\end{figure}

 In Fig.~\ref{results} we show $\sigma$ as a function of $\beta$ for
two choices of $\alpha$, representative of the two types of
regime. In both cases we set $a=1$, $b=1$ and $\Gamma=0.01$, and
we vary the quench rate as $\tau_Q=2^n$,  $n=1,2,..,9$. The final
defect densities are obtained by averaging over ten independent
realizations. Production runs were performed on periodic lattices
with $16000$ points and space and time steps of $\delta x=0.125$
and $\delta t=0.1$.  We found that coarse-graining $\phi$
 over eight lattice points was enough to eliminate any unwanted
fluctuations of the field.
The error-bars, not shown in the plot for
clarity, were obtained by calculating the standard deviation over
the ensemble of $10$ sample runs, and are of order of $10\%$ of
the values observed for $\sigma$.

Starting with the high-dissipation results with $\alpha=1.0$, we
see that within error-bars, the scaling exponent is approximately
constant for $\beta<0.5$. The value measured $\sigma=0.28$,
coincides with that obtained for similar simulation parameters in
the case of a single field system \cite{us}. One should note that
the method used to determine the final defect density is known to
lead to a slight over-estimate in $\sigma$ when compared to other,
computationally more demanding, approaches (for a discussion see
\cite{laguna}). This leads to
 the deviation from the theoretical
prediction $\sigma\sim1/4$ for the over-damped regime. With this caveat
in mind, and remembering that the transition between the under and
over-damped regimes is
continuous, we will proceed as in \cite{us} and use the value $\sigma\sim0.30$
to distinguish between the two types of behaviour.
 In the following, all values of
$\sigma$ should be taken in reference to those obtained in the low-$\beta$
parameter region.
 As is clear in Fig.~\ref{results}, as $\beta$ increases, the scaling exponent
goes down signaling a departure from the canonical KZ-scaling behaviour, 
moving deeper into the dissipative regime. For
values of $\beta$ in the vicinity of $0.80$, $\sigma$ reaches a
stable plateau, remaining constant (within error-bars) up to
$\beta\sim 0.97$ when a sharp increase takes place.
 The results for the relativistic case $\alpha=0.6$ show a very
similar pattern, with $\sigma$ steadily decreasing towards a flat regime,
followed by a sudden rise as $\beta\rightarrow 1$.

 The initial period of slow decrease in $\sigma$ for values of $\beta$
up to $0.7-0.8$ can be related to the fact that the
effective physical consequences of the dissipation should be measured in
terms of the relevant masses of the problem. The relaxation time
scale that determines the type of behaviour (dissipative {\it vs}
relativistic) is given by the inverse of $\alpha^2/m$, where $m$ stands
for the lowest mass in the theory. As $\beta$ increases and $T<T_c$,
the mass of one of the excitations of the vacuum approaches zero. When
$\beta=1$ and the original Lagrangian becomes explicitly $U(1)$
symmetric, this degree of freedom corresponds to the massless
Goldstone boson. More specifically, the masses of the vacuum
excitations for the potential in Eq.~(\ref{potential}) are given
by $m_1^2=-2 a$ and $m_2^2=-a (1-\beta)$. Clearly, $m_2$ takes
increasingly small values as $\beta$ approaches $1$. This leads to
a rise in the largest effective dissipation scale
$\alpha^2/m_2$, shifting the system towards the over-damped limit and
decreasing the value of $\sigma$. In order to check this we simulated a one
field relativistic system with a varying mass term. We observed as
expected, that as the mass was reduced the value of $\sigma$
decreased, with the system displaying dissipative effects.

 How is this trend towards lower values of the scaling exponent
stopped? To understand the overall behaviour of $\sigma$ as
displayed is Fig.~\ref{results} one must keep in mind that as
$\beta$ takes larger values, the interaction between kinks and
anti-kinks changes considerably. As discussed in Section
\ref{interaction}, for large separations, there is a repulsive
interaction between kinks and anti-kinks with the same charge.
Note that the effect of the repulsive interaction should not be
felt immediately after $\beta>0.5$, as the height of the potential
barrier is initially very small. Nevertheless, since the energy
needed to overcome the repulsive barrier increases with $\beta$,
we expect kink/anti-kink annihilation to be inhibited for larger
values of this parameter. This should lead to survival of a larger
number of  kink pairs, increasing the final number of defects.
What would be the signature of this mechanism in the scaling of
the final defect density with quench rate? Clearly the effect
would be felt more strongly for fast quenches, where defect
densities are higher and annihilation is likely to play a bigger
role. An increase in kink survival rates for low values of
$\tau_Q$ should lead to a steeper distribution $\rho(\tau_Q)$,
that is, larger values of the exponent $\sigma$.

\begin{figure}
\scalebox{0.45}{\includegraphics{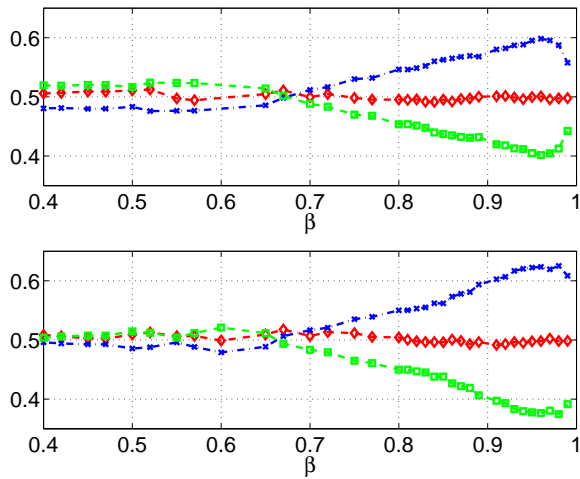}} \caption{
Fraction of same-charge kink/anti-kink pairs (crosses) and
opposite charged pairs (squares) as a function of $\beta$. For
reference we also show the fraction of negative charge kinks
(diamonds). The bottom plot corresponds to the under-damped regime
$\alpha=0.6$ and the top plot to the over-damped case $\alpha=1$.}
\label{pairs}
\end{figure}

 The mechanism above provides an explanation to why the decrease in $\sigma$
 is halted for values of $\beta$ roughly halfway between $0.5$ and
 $1.0$. In order to confirm this scenario we identified neighbouring pairs
 of kinks and anti-kinks in the simulation and determined their charges
 by looking at the sign of $\eta$ in their cores. Using the knowledge
 of the spatial distribution of charges at the final time of each simulation,
 we counted the number of defect pairs with equal and opposite charge
 respectively.
 If same-charge pairs are being prevented from annihilating, then their
 numbers should be in excess of those of opposite charged pairs.
 In Fig.~\ref{pairs} we compare the total number ({\it i.e.} summed
 over all quenches) of both types of kink/anti-kink pairs as a function
 of $\beta$, in the dissipative and relativistic cases.  The results
 match our expectations well, with the fraction of pairs with equal
 charge deviating little from $0.5$ up to values of $\beta\sim 0.7$.
 Above that threshold there is a decrease in the percentage of pairs
 with attracting charges, signaling the survival of equal charge pairs
 due to the rise in the repulsive potential barrier. We also confirmed that
 this effect is more marked for fast quenches, as discussed above.
 For reference,
 the plots include the data of the fraction of pairs with negative charge
 as well. As expected, as a consequence of
 the symmetry of the theory, this quantity remains equal to $0.5$
 within error-bars throughout the whole range of $\beta$.

 Finally we will focus on the apparently anomalous behaviour of the
 system for values of  $\beta$ very near $1$, in which limit there is no topological charge.
 For the three points
 with highest $\beta$, {\it i.e.} for $\beta > 0.96$, there is a marked
 increase in the exponent $\sigma$, accompanied by a decrease in the fraction
 of same-charge pairs.  A closer look at the data reveals that the change in
 relative amount of types of kink/anti-kink pairs is caused exclusively by the
 number of pairs with opposite charge going up by a considerable amount.
 We note that the results for these three values of $\beta$ should be taken
 with care since we are in a regime where the kink core size becomes
 increasingly large. In particular, for fast quenches with high final
 defect densities, the defect-defect separation becomes of the order
 of magnitude of the core size. Taking the most extreme case $\beta=0.99$,
 we find for the simulation parameters $l=10$, which implies that each
 defect has a spatial extension of the order of $20$. Our simulation
 box should in principle not be able to accommodate more than $100$ of
 these kinks, and for fast quenches the number observed is indeed very close
 (for example $80$ in the relativistic case, for the smallest $\tau_Q$).
 In other words, the correlation
 length leading to domain formation at the transition is close to the
 core kink size. These cases should be understood as being effectively in the
 $U(1)$ regime, with winding springs being formed with no
 spatially localized core. The dynamics of these objects
 should be described in terms of unwinding processes and this might lead to a
 slower annihilation rate than traditional kink/anti-kink collisions.
 In terms of the interaction potentials in Fig.~\ref{attraction} this
 corresponds to the flattening of the attraction potential as $\beta\rightarrow 1$. A study of the time-scales involved in the departure from equilibrium
 in such case (using for example the techniques developed in \cite{Gleiser}) 
 could also shed some light on the dynamics of these systems.
 A detailed analysis of the unwinding process
 and its role in defect production in a phase-transition will
 be the focus of a future publication dedicated exclusively to the $U(1)$
 theory in $1+1$ dimensions \cite{future}.

\section{Conclusions}

 We studied kink formation in 1+1 dimensions, in a scalar
 theory where an added field can lead to condensation in the
 defect core. We obtained explicit expressions for the non-trivial
 defect profiles, and determined the region of parameter space
 for which these are stable. This introduces an extra degree of
 complexity in the theory as kinks acquire a charge defined in terms
 of the sign of the core condensate.
 As a consequence, the process of defect production at a non-equilibrium
 phase transition suffers a qualitative change, with annihilation of kinks
 and anti-kinks being suppressed due to the repulsive nature of their
 interaction. The degree to which this takes place depends on
 the parameters of the potential. As a consequence we observed deviations
 from traditional Kibble-Zurek scaling, with the value of the
 scaling parameter showing a clear, measurable signature of this effect.

 These results open interesting possibilities for the case of higher
 dimensional theories. Though there seems to be good evidence by now in favour
 of the stability of vortons in  $U(1)\times U(1)$ theories in 3+1 dimensions
 \cite{Shellard}, the question of formation still remains unresolved.
 Vortons are very fragile objects making a numerical study of their
 production in a phase-transition impossible, the size of the
 domains required being severely constrained by computational limitations.
 The model described in this article can be generalized to 2+1 dimensions
 and an extra condensate field introduced. Such $Z_2 \times U(1)$ theory
 would display domain walls formed by kinks of the $Z_2$ field,
 with the $U(1)$ field condensing in their cores and leading to vorton-like
 solutions. The lower space-time dimensionality of this model
 could be make numerical simulations of vorton formation possible - a line
 of research we will be following in a future publication.

\section{Acknowledgements}

 N. D. Antunes was funded by PPARC. P. Gandra was supported by FCT,
 grant number PRAXIS XXI BD/18432/98. RR would like to acknowledge support
 from the ESF COSLAB programme.

%\section*{References}

\end{document}